\documentstyle[psfig,12pt]{article}
\def\etal{{\it et al.}}

\def\beq{\begin{equation}}
\def\eeq{\end{equation}}
\def\bea{\begin{eqnarray}}
\def\eea{\end{eqnarray}}
\def\bq{\begin{quote}}
\def\eq{\end{quote}}

\def\bq{\begin{quote}}
\def\eq{\end{quote}}

\def\bq{\begin{quote}}
\def\eq{\end{quote}}

\def\gappeq{\mathrel{\rlap {\raise.5ex\hbox{$>$}}
{\lower.5ex\hbox{$\sim$}}}}

\def\lappeq{\mathrel{\rlap{\raise.5ex\hbox{$<$}}
{\lower.5ex\hbox{$\sim$}}}}

\def\bbz{fa Z \kern-8.9pt Z}

\begin{document}

\baselineskip 18pt
\newcommand{\sheptitle}
{Radiative Corrections to Chargino Production
in Electron-Positron Collisions with Polarised Beams}

\newcommand{\shepauthor}
{Marco A. D\'\i az$^{1}$, Stephen F. King${^2}$ and Douglas A.
Ross${^2}$}

\newcommand{\shepaddress}
{${^1}$Departmento de F\'\i sica,
Universidad Cat\'olica de Chile,
Av. Vicu\~na Mackenna 4860, 
Santiago, Chile
\\
$^{2}$Department of Physics
and Astronomy, University of Southampton, Southampton, SO17 1BJ,
U.K.}

\newcommand{\shepabstract}
{We study radiative corrections to chargino production
at linear colliders with polarised electron beams.
We calculate the one-loop corrected cross-sections 
for polarised electon beams due to three families
of quarks and squarks, working in the $\overline{MS}$ scheme, extending our
previous calculation of the unpolarised cross-section with one-loop
corrections due to the third family of quarks and squarks.
In some cases we find rather large corrections to the tree-level
cross-sections. For example for the case of right-handed polarised electrons
and large $\tan \beta$ the corrections can be
of order 30\%, allowing sensitivity to the squark mass
parameters.}

\begin{titlepage}
\begin{flushright}
hep-ph/0008117\\
UCCHEP/13-00
\end{flushright}
\begin{center}
{\large{\bf \sheptitle}}
\bigskip \\ \shepauthor \\ \mbox{} \\ {\it \shepaddress} \\ \vspace{.5in}
{\bf Abstract} \bigskip \end{center} \setcounter{page}{0}
\shepabstract
\end{titlepage}

Charginos are important in Supersymmetry (SUSY) for several
reasons. To begin with they will possibly be the next-to-lightest
SUSY particles (after the lightest neutralino) and so be amongst
the first supersymmetric particles to be discovered. Secondly,
being colour singlets, they provide a clean laboratory for
studying and extracting the fundamental parameters of SUSY.
Thirdly they are naturally produced in a polarised state, and
their polarisation is imprinted onto the angular distribution of
their decay products, enabling important information about the
nature of the underlying SUSY theory to be extracted from
experiment.

Since LEP has failed to find evidence for any SUSY particles \cite{experim}, 
one must await the construction of the next generation of
electron-positron machines, which will be linear colliders, to
perform high precision studies of chargino physics. Although
charginos may be discovered earlier at hadron colliders, it is
only at such linear colliders, with the added advantage of
polarised beams, that the parameters of SUSY can begin to be
extracted with any precision \cite{unpolar}. In the context of such 
high energy, high precision colliders, the polarisation  properties of the
produced charginos can be studied via the angular distribution of
the decay products as has been recently discussed by several
groups \cite{polarized}.

Such studies involve helicity amplitudes for both chargino
production and decay which undergo quantum interference due to the
short lifetime of the charginos. Thus far both the production and
decay helicity amplitudes have only been studied to lowest order,
although the spin averaged cross-section for chargino production
in $e^+e^-$ collisions has been calculated at one-loop, including
third family quark and squark loop corrections 
\cite{DKR,oneloop}, and the
radiative corrections to the chargino self-energy has been
calculated including all one-loop radiative corrections
\cite{pierce}.

In this letter, then, we present the first study of the radiative
corrections to chargino production in
electron-positron collisions, including contributions from 
three generations of squark
and quark loops, for the case of polarised electron beams.
Our main purpose here is to study such corrections numerically,
and show that the effects may be rather large in some cases.
Although the radiative corrections in the cross-sections 
are of order 1-10\% in general, for the cross-section
for right-handed electrons 
we observe strong cancellations in the tree-level result
due to interference terms with negative signs, 
and in this case the radiative corrections may be of order
30\% for large $\tan \beta$.

We consider pair production of charginos with momenta $k_1$ and
$k_2$ in electron-positron scattering with incoming momenta $p_1$
and $p_2$:
\beq e^+(p_2)+e^-(p_1)\rightarrow
\tilde{\chi}^+_b(k_2)+\tilde{\chi}^-_a(k_1) 
\eeq
where we take $\tilde{\chi}^+$ to be the particle and
$\tilde{\chi}^-$ to be the antiparticle, with the Feynman rules as
given in Haber and Kane \cite{HK}. 
Henceforth we drop the subscripts
$b$ and $a$, but understand that the two charginos have masses
$m_b$ and $m_a$ respectively and in general $m_b \neq m_a$.

In \cite{choi} it was shown that at the tree-level one can (after 
appropriate Fierz transformation) write the scattering amplitudes as
\begin{eqnarray} & & 
\frac{-ie^2}{s}        
    \left[ \bar{v}(e^+) \gamma^\mu \frac{(1-\gamma^5)}{2} u(e^-) \right]
\nonumber \\ & & 
\left( Q_{LL}^{(0)}\left[  \bar{u}(\tilde{\chi}^+)
  \gamma_\mu \frac{(1-\gamma^5)}{2} v(\tilde{\chi}^-) \right]
  + Q_{LR}^{(0)}
   \left[\bar{u}(\tilde{\chi}^+)
    \gamma_\mu \frac{(1+\gamma^5)}{2} v(\tilde{\chi}^-) \right] \right)
  \label{kal1} \end{eqnarray}
for left-polarized incident electrons, with a similar result
for right-polarized incident electrons with the electron projection
operator $\frac{(1-\gamma^5)}{2}$ replaced by $\frac{(1+\gamma^5)}{2}$
and $Q_{L\beta}^{(0)}$
replaced by $Q_{R\beta}^{(0)}$, where the superscript zero indicates
tree-level. Note that since we take the
positive chargino to be the particle, the
index $\beta = L,R$ is related to that in \cite{choi} by 
$L \leftrightarrow R$.

At one-loop order we find a more general structure. Nevertheless the 
amplitudes may be written as
\beq 
\frac{-ie^2}{s}        
    \left[ \bar{v}(e^+) \gamma^\mu \frac{(1 - \gamma^5)}{2} u(e^-) \right]
    \sum_{i=1\cdots 5}  Q_{Li}^{(1)}  \left[ \bar{u}(\tilde{\chi}^+) \Gamma^i
    v(\tilde{\chi}^-) \right], 
\label{oneloop}
\eeq
for left-polarized incident electrons, with a similar result
for right-polarized incident electrons
with the electron projection
operator $\frac{(1-\gamma^5)}{2}$ replaced by
$\frac{(1+\gamma^5)}{2}$, 
and $Q_{Li}^{(1)}$
replaced by $Q_{Ri}^{(1)}$
where the superscript unity
indicates one-loop, and where 
\begin{eqnarray}
\Gamma^1 & = & \frac{(1+\gamma^5)}{2} \\
\Gamma^2 & = & \frac{(1-\gamma^5)}{2} \\
\Gamma^3 & = & \gamma^\nu\frac{(1+\gamma^5)}{2} \\
\Gamma^4 & = & \gamma^\nu\frac{(1-\gamma^5)}{2} \\
\Gamma^5 & = & \sigma^{\nu\rho} \ = \ \frac{i}{2} 
\left[ \gamma^\nu,\gamma^\rho\right]  \end{eqnarray}
Note that the coefficients $Q_{L1}^{(1)}$ and $Q_{L2}^{(1)}$ are vectors,
$Q_{L3}^{(1)}$ and $Q_{L4}^{(1)}$ are two-rank tensors, and $Q_{L5}^{(1)}$
is a three-rank tensor, and similarly for $Q_{Ri}^{(1)}$.

\begin{figure}
\centerline{\protect\hbox{\psfig{file=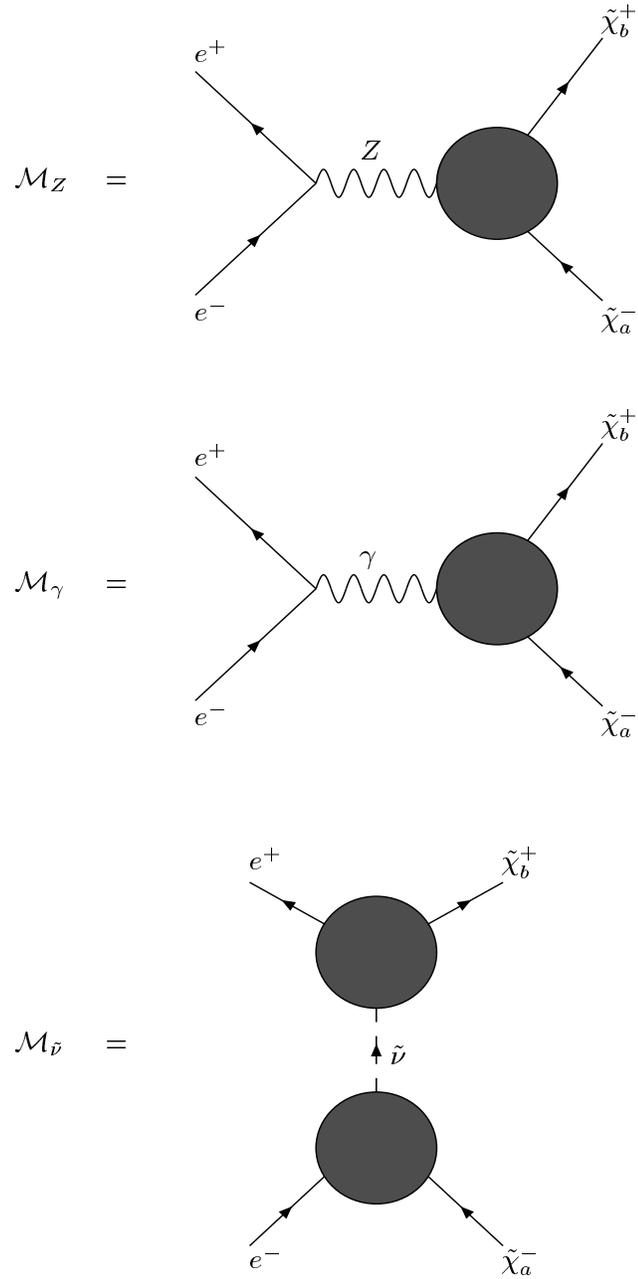,height=17cm,width=0.55
\textwidth}}}
\caption{One--loop renormalized $M_Z$, $M_{\gamma}$ and 
$M_{\tilde\nu_e}$ amplitudes in the approximation where three families of
quarks and squarks are considered inside the loops.} 
\label{ZGSneu1lAmplitudes} 
\end{figure} 
In the presence of one-loop corrections,
due to the three families of quarks and squarks,
the amplitude for $e^+e^-\rightarrow \tilde{\chi}^+_b\tilde{\chi}^-_a$
may be expressed as the sum of three amplitudes $M_Z$, $M_{\gamma}$,  
$M_{\tilde{\nu}}$ as shown in Fig.~\ref{ZGSneu1lAmplitudes}. 
The shaded bubbles
in that figure are one--loop renormalized total vertex functions defined as
$i{\cal G}_{Z\chi\chi}^{ab}$, $i{\cal G}_{\gamma\chi\chi}^{ab}$, 
$i{\cal G}_{\tilde{\nu_e}e\chi}^{+b}$, and
$i{\cal G}_{\tilde{\nu_e}e\chi}^{-a}$.
In the total vertex functions we include the tree level vertex, the
one--particle irreducible vertex diagrams plus the vertex counterterm, and
the one--particle reducible vertex diagrams plus their counterterms.
Although the detailed expressions for the
total vertex functions is quite complicated, by exploiting the possible
Lorentz structures of the diagrams it is possible to
express them in terms of just a few form factors which are
generalisations of those presented for the case
of the third quark and squark family in \cite{DKR}, where explicit 
expressions may be found.
These form factors
may in turn be related to the quantities $Q_{L i}^{(1)}$
defined in Eq.~(\ref{oneloop}), and similarly for $Q_{R i}^{(1)}$,
as we shall show in detail
in a forthcoming publication \cite{DKR2}.
\footnote{The one-loop corrections due to quarks and squarks
considered here do not involve the operator $\Gamma^5$.}

One of the main purposes of this letter is to 
examine the numerical effect of these corrections, and show
that in some cases they may be rather large.
For definiteness 
at the $\overline{MS}$ scale $Q=M_Z$ we take the $SU(2)_L$ gaugino mass
$M_2$ to be 165 GeV, 
the $\mu$ parameter to be 400 GeV and the remaining trilinear $A$
and (degenerate) squark soft mass parameters
$A=M_Q=M_U=M_D$ to be 500 GeV initially.
We also assume a sneutrino mass of 500 GeV.
Note that the lighter chargino (1) will be mainly wino,
with a mass $\approx M_2$, and the heavier chargino (2) will
be mainly higgsino with a mass $\approx \mu$ 
in this example.

\begin{figure}
\centerline{\protect\hbox{\psfig{file=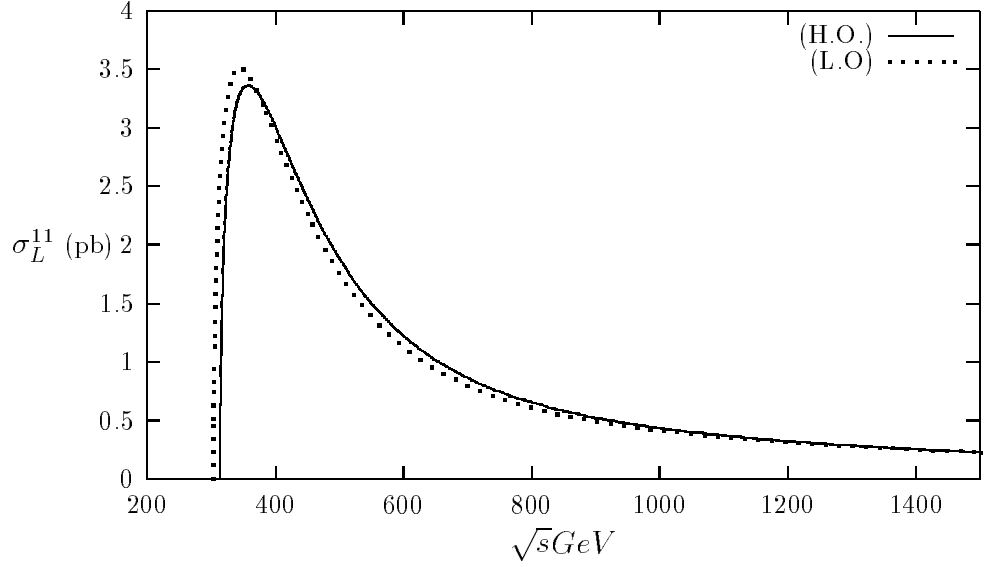,height=7.5cm,width=0.85
\textwidth}}}
\caption{Lowest order (L.O.) and higher order (H.O.)
cross-section for lightest chargino pair production
for left-polarized electrons with $\tan \beta =5$ and the
other parameters as given in the text.} 
\label{L5} 
\end{figure}

Assuming these parameters with $\tan \beta=5$,
Figure \ref{L5} shows the cross-section for the production
of the lightest chargino pair for beams of
left-handed polarized electrons as a function of the
centre of mass energy. 
The effect of radiative corrections
is to reduce the cross-section by a few per cent,
with a noticeable shift in the lightest chargino mass
threshold due to the more steeply rising threshold.

\begin{figure}
\centerline{\protect\hbox{\psfig{file=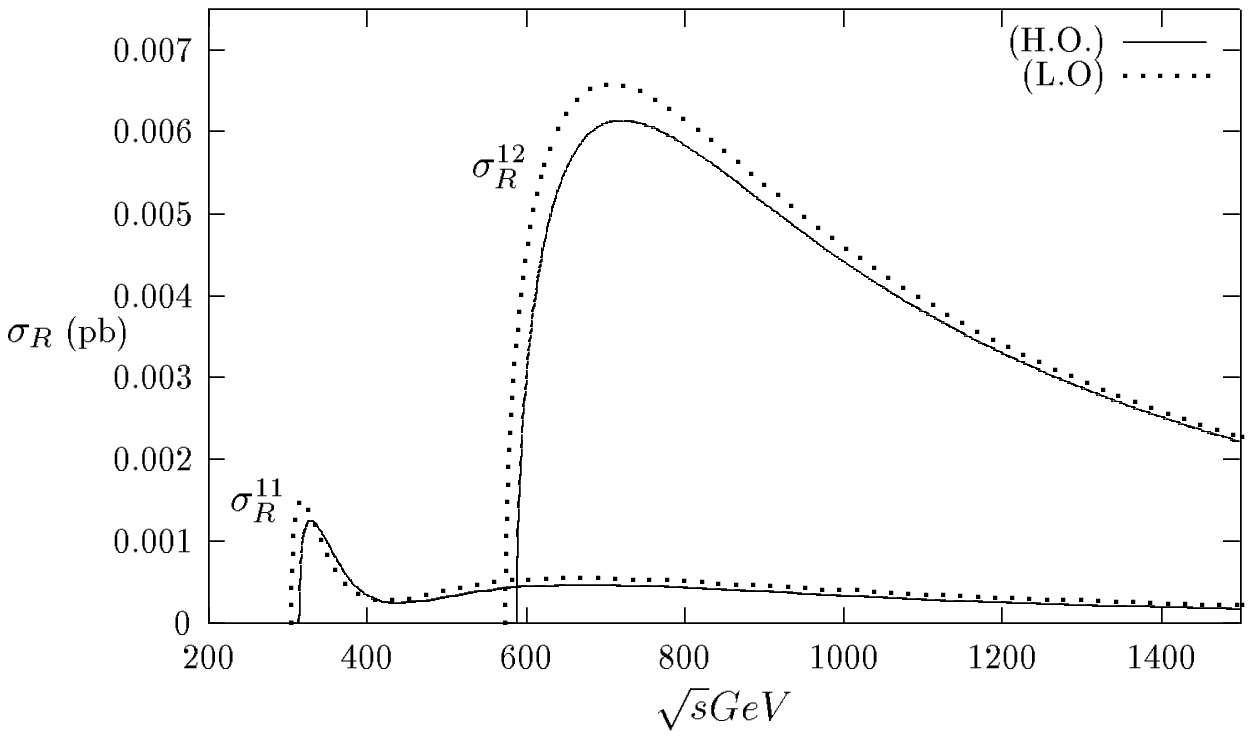,height=7.5cm,width=0.85
\textwidth}}}
\caption{Lowest order (L.O.) and higher order (H.O.)
cross-sections for lightest and unequal mass chargino
pair production for
right-polarized electrons with $\tan \beta =5$ and the
other parameters as given in the text.} 
\label{R5} 
\end{figure} 
\begin{figure}
\centerline{\protect\hbox{\psfig{file=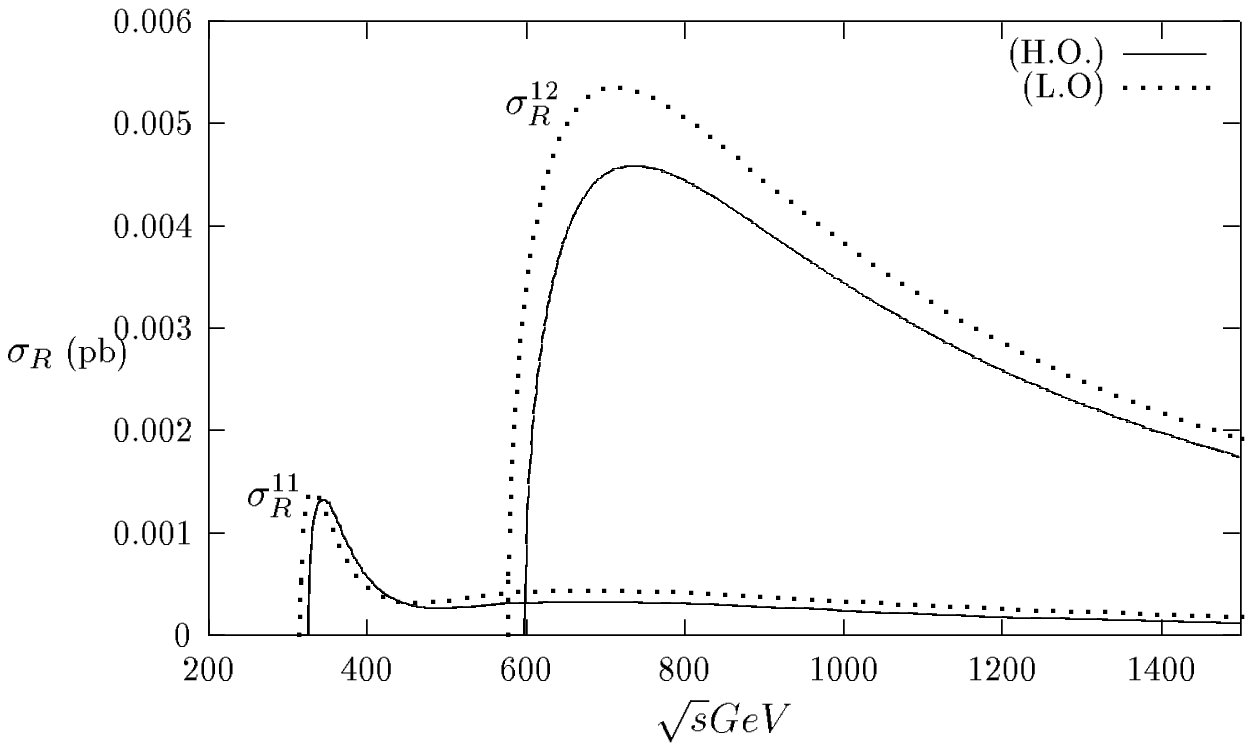,height=7.5cm,width=0.85
\textwidth}}}
\caption{Lowest order (L.O.) and higher order (H.O.)
cross-sections for lightest and unequal mass chargino
pair production for
right-polarized electrons with $\tan \beta =50$ and the
other parameters as given in the text.} 
\label{R50} 
\end{figure} 

Figure \ref{R5} displays the cross-section for the production
of both lightest and unequal mass chargino pairs for beams of
right-handed polarized electrons as a function of the
centre of mass energy, for the same parameters as before with 
$\tan \beta =5$. 
The unequal mass cross-section $\sigma^{12}$ refers to 
$b=1,a=2$, which is equal to the cross-section for $b=2,a=1$
assuming CP to be conserved, although the two cross-sections are not
added together in the figures.
Note that the cross-section
for the lightest chargino pairs with right-handed electrons in Figure
\ref{R5} are 
about 500 times smaller than with left-handed electrons in Figure \ref{L5},
nevertheless with an integrated luminosity of $10^6 \ pb^{-1}$
it will be easily measurable. 
\footnote{
The reason for the smallness of
the cross-section for lightest chargino production with
right-handed electrons
is due to a destructive interference between the photon and
$Z$ diagrams, 
compared to a constructive interference with left-handed electrons.
This is due to the approximately axial couplings of electrons to the $Z$.
The absence of the sneutrino
exchange diagram for right-handed incident electrons
then guarantees a small cross-section in this case.
For the production of unequal mass charginos and right-handed electrons
the photon exchange
diagram is not present (at least at tree-level)
and the cancellation does not occur,
leading to the larger cross-section than the equal mass case
in Figure \ref{R5}.}
The radiative corrections involving right-handed 
incident electrons in Figure \ref{R5} are larger than for left-handed
incident electrons in Figure \ref{L5}, and may now be as large as
about 10\%. Note the shift in the second chargino mass threshold.

Increasing $\tan \beta$ to 50 makes very little difference
to the tree-level and one-loop corrected cross-section
for left-handed electrons, as compared to the results
for $\tan \beta =5$ in Figure \ref{L5}.
However for right-handed electrons, increasing
$\tan \beta$ to 50 leads to the much larger radiative
corrections shown in Figure \ref{R50} as compared to 
Figure \ref{R5}. 
In view of the large radiative corrections in this
case, we proceed to study these regions in a little more detail.

\begin{figure}
\centerline{\protect\hbox
{\psfig{file=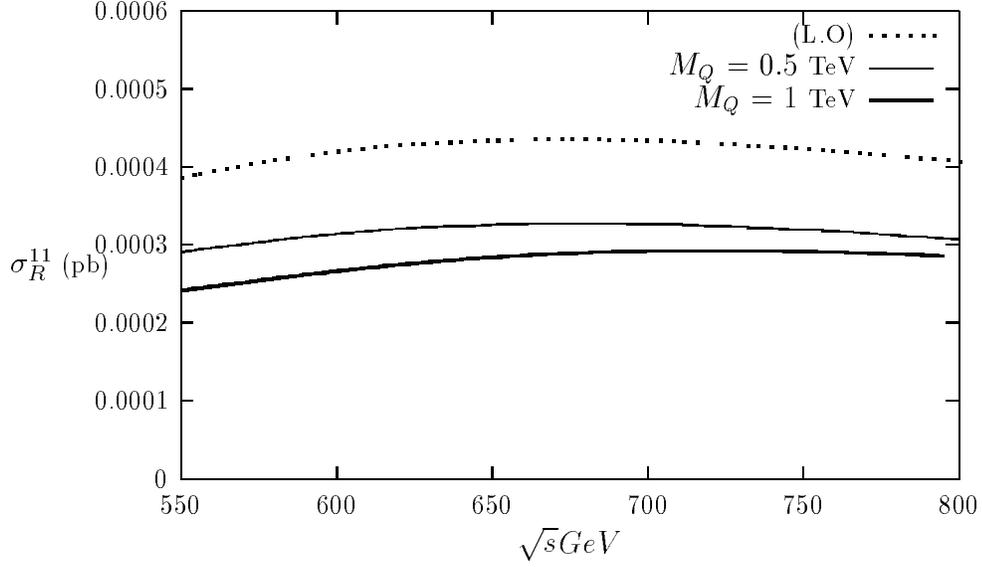,height=7.5cm,width=0.85
\textwidth}}}
\caption{Detailed blow-up of
cross-sections for right-polarized electrons
for the lightest chargino pair with $\tan \beta =50$.
The H.O. cross-sections are for degenerate squark soft mass
parameters of 0.5 TeV and 1 TeV.} 
\label{R50blowup11} 
\end{figure} 
\begin{figure}
\centerline{\protect\hbox
{\psfig{file=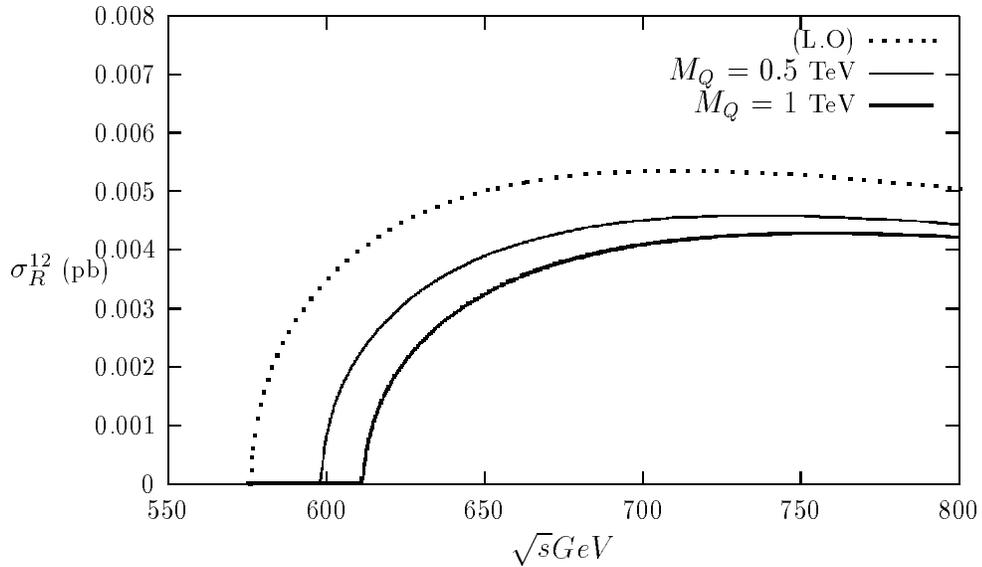,height=7.5cm,width=0.85
\textwidth}}}
\caption{Detailed blow-up of the
cross-sections for right-polarized electrons
for the unequal mass chargino pair with $\tan \beta =50$.
The H.O. cross-sections are for degenerate squark soft mass
parameters of 0.5 TeV and 1 TeV.} 
\label{R50blowup12} 
\end{figure} 

In Figure \ref{R50blowup11} we magnify a region of 
the cross-section for lightest chargino pair production
for right-handed electrons above the threshold region
in Figure \ref{R50}.
As already remarked, the effect of radiative corrections is 
quite large. The corrections
depend on the (degenerate) squark mass as shown in
Figure \ref{R50blowup11}, where squark masses of
500 GeV (1 TeV) leads to negative corrections of about 25\% (35\%).

In Figure \ref{R50blowup12} we magnify a region of the cross-section
for unequal mass chargino production for right-handed electrons,
this time concentrating on the threshold region of Figure \ref{R50}
where the corrections appear to be largest.
The shift in the second chargino mass is clearly apparent,
and is about 20 GeV (30 GeV) for squark masses of 500 GeV (1 TeV).
The peak cross-section is reduced by about 20\% (30\%) 
for squark masses of 500 GeV (1 TeV).
The sensitivity of the higher order corrections to the squark soft mass
parameters for the case
of right-handed incident electrons, shown in Figures \ref{R50blowup11}
and \ref{R50blowup12}, means that for the case of large $\tan \beta$
at least, information about the soft squark masses 
may be inferred from
sufficiently accurate measurements of the chargino production
cross-sections.

In summary we have seen that for the case of right-handed
polarised electron beams the effects of radiative corrections
due to loops of quarks and squarks may give significant
corrections to the lowest order result, particularly for
large values of $\tan \beta$.
These results highlight the importance of being able to 
measure cross-sections with polarised electron beams at
future linear colliders.
Such large radiative corrections must be taken
into account if the underlying SUSY parameters are to be accurately
extracted from the experimentally measured chargino cross-sections.
The corrections will involve the squark masses which may therefore
be probed via radiative corrections.
The effect of radiative corrections on the production
of polarised charginos will be considered in a future publication
\cite{DKR2}.

\noindent{\bf Acknowledgements:}
We are grateful to the Royal Society and CONICYT for
a joint grant (No.~1999-2-02-149) which made this research possible.
M.A.D. was also supported by CONICYT grant No.~1000539.

\end{document}